\documentclass[aps,prd,twocolumn,superscriptaddress,preprintnumbers,floatfix,nofootinbib,notitlepage]{revtex4-1}

\usepackage{graphicx}
\usepackage{amsmath,amsfonts,amssymb}
\usepackage{hyperref}
\usepackage{color}

\def\lsim{\mathrel{\raise.3ex\hbox{$<$\kern-.75em\lower1ex\hbox{$\sim$}}}}
\def\gsim{\mathrel{\raise.3ex\hbox{$>$\kern-.75em\lower1ex\hbox{$\sim$}}}}

\newcommand{\be}{\begin{equation}}
\newcommand{\ee}{\end{equation}}
\newcommand{\bea}{\begin{equation}\begin{aligned}}
\newcommand{\eea}{\end{aligned}\end{equation}}
\newcommand{\td}{{\rm d}}

\newcommand{\khat}{\hat{\boldsymbol{k}}}

\begin{document}

\preprint{KCL-PH-TH/2019-88}

\title{On bubble collisions in strongly supercooled phase transitions}

\author{Marek Lewicki}
\email{marek.lewicki@kcl.ac.uk}
\affiliation{Physics Department, King's College London, London WC2R 2LS, UK}
\affiliation{Faculty of Physics, University of Warsaw ul.\ Pasteura 5, 02-093 Warsaw, Poland}
\author{Ville Vaskonen}
\email{ville.vaskonen@kcl.ac.uk}
\affiliation{Physics Department, King's College London, London WC2R 2LS, UK}
\affiliation{NICPB, R\"avala 10, 10143 Tallinn, Estonia}

\begin{abstract}
We study strongly supercooled cosmological phase transitions. We perform numerical lattice simulations of two-bubble collisions and demonstrate that, depending on the scalar potential, in the collision the field can either bounce to a false vacuum or remain oscillating around the true vacuum. We study if these cases can be distinguished from their gravitational wave signals and discuss the possibility of black hole formation in the bubble collisions.
\end{abstract}

\maketitle

\section{Introduction}

Various particle physics models include symmetries that are broken at low energies. At high temperatures in the early Universe, these symmetries are restored by thermal effects. Symmetry breaking phase transitions happen as the Universe expands and cools down. Some of these phase transitions may be of first-order, and therefore generate a gravitational wave (GW) background~\cite{Witten:1984rs}. If this background is sufficiently strong, it can be probed by future GW observatories, such as LISA~\cite{2017arXiv170200786A}. 

A first-order phase transition proceeds by nucleation of bubbles of a new energetically favoured vacuum state~\cite{Coleman:1977py,Callan:1977pt,Linde:1981zj}. These bubbles then expand, eventually collide with each other and finally turn the whole universe in the true vacuum state. It is possible that the vacuum energy of the false vacuum surpasses the thermal energy before the transition finishes. In this case the bubble walls accelerate almost to the speed of light and accumulate a large amount of energy before their collisions.

The friction caused by the thermal plasma can slow down the wall such that it reaches a terminal velocity~\cite{Bodeker:2009qy,Bodeker:2017cim}. Most of the energy released in the transition is then in the plasma motions and the GW signal from the phase transition is dominated by sound waves and turbulence in the plasma~\cite{Kamionkowski:1993fg,Caprini:2009yp,Hindmarsh:2013xza,Cutting:2019zws}. Models where this in general happens include many extensions of the Standard Model related for example to electroweak baryogenesis~\cite{No:2011fi,Vaskonen:2016yiu,Dorsch:2016nrg,Artymowski:2016tme,Beniwal:2017eik,Beniwal:2018hyi} or hidden/dark sectors~\cite{Espinosa:2008kw,Baldes:2018emh,Breitbach:2018ddu,Fairbairn:2019xog}. However, if the transition is sufficiently strongly supercooled, the terminal velocity is not reached before the bubble walls collide~\cite{Ellis:2019oqb}. In this case the GW signal is dominantly sourced by the scalar field gradients~\cite{Kosowsky:1992vn,Cutting:2018tjt}. Such strongly supercooled phase transitions are expected especially in models that are classically scale invariant~\cite{Randall:2006py,Konstandin:2011dr,Jinno:2016knw,Iso:2017uuu,Marzola:2017jzl,vonHarling:2017yew,Marzo:2018nov,Baratella:2018pxi}. 

In this paper we study the latter case by performing numerical lattice simulations of two-bubble collisions. We begin in Sec.~\ref{sec:theory} by reviewing the theory of bubble nucleation and growth. Then, in Sec.~\ref{sec:simulations} we present the results of our simulations and show that depending on the scalar potential, the field can either remain oscillating around the true vacuum or bounce to the false vacuum immediately after the collision. In Sec.~\ref{sec:gws} we study the GW signal from the phase transition in both of these cases. Our results indicate that the differences in the GW signal may be too small in order to distinguish these cases.  This conclusion may, however, change when multiple bubbles are considered. Finally in Sec.~\ref{sec:bhs} we discuss the formation of black holes in the bubble collisions. In particular, our results indicate that the main assumption in Ref.~\cite{Khlopov:1998nm}, that the false vacuum bubble formed in the collision becomes spherical, is wrong and therefore we conclude that their mechanism for black hole formation in bubble collisions does not work.

\section{Bubble nucleation and growth}
\label{sec:theory}

Let us consider a scalar field that is initially in a local minimum at $\phi=0$ of its potential $V(\phi)$ but the global minimum of $V(\phi)$ is at $\phi\neq 0$ (see Fig.~\ref{fig:pot}). The barrier tunneling probability to the energetically favoured phase at $\phi\neq 0$ is $\Gamma\propto e^{-S_E}$~\cite{Coleman:1977py,Callan:1977pt,Linde:1981zj}, where $S_E$ is the Euclidean action of the scalar field,
\be
S_E = \int \td^4 x \left[\frac{1}{2}(\partial_t \phi)^2 + \frac{1}{2}(\nabla \phi)^2 + V(\phi)\right] \,.
\ee
The tunneling probability is dominated by the path that minimizes the action. This is given by the $O(4)$ symmetric classical configuration that is solution of
\be \label{eq:O4initialbubble}
\partial_r^2 \phi + \frac{3}{r} \partial_r \phi = V'(\phi) \,,
\ee
where $r^2 = t^2 + x^2 + y^2 + z^2$, with boundary conditions $\partial_r\phi = 0$ at $r=0$ and $\phi\to 0$ at $r\to \infty$. Such a solution represents a bubble inside of which the field is near the true vacuum while outside it goes to the false vacuum background.

At non-zero temperature, $T\neq 0$, the situation can change significantly. As discussed in Ref.~\cite{Linde:1981zj}, at $T\gg r_0$, where $r_0$ is the radius of the $O(4)$ symmetric solution, that can be roughly estimated as the radius that maximizes $|\partial_r \phi|$, the action reduces to $S_E \simeq S_3/T$, where ($r^2=x^2+y^2+z^2$)
\be
S_3 = 4\pi \int r^2 \td r\left[ \frac{1}{2} (\partial_r \phi)^2 + V(\phi) \right] \,.
\ee
The corresponding equation of motion is
\be \label{eq:O3initialbubble}
\partial_r^2 \phi + \frac{2}{r} \partial_r \phi = V'(\phi) \,,
\ee
which again should be solved with boundary conditions $\partial_r\phi = 0$ at $r=0$ and $\phi\to 0$ at $r\to \infty$. The bubble in this case is in general smaller than in the $T\ll r_0$ (or $O(4)$ symmetric) case.

Next, let us describe the evolution of the bubbles after nucleation. In Minkowski space ($g_{\mu\nu} = {\rm diag}(1,-1,-1,-1)$) the scalar field action is
\be \label{eq:SM}
S = \int \td^4 x \left[\frac{1}{2}(\partial_\mu \phi)^2 - V(\phi)\right] \,,
\ee
and the time evolution of $\phi$ is governed by the Klein-Gordon equation
\be  
\partial_\mu \partial^\mu \phi = -V'(\phi) \,.
\ee

Consider now an $O(3)$ symmetric bubble nucleation. It is convenient to describe the evolution of this field configuration in spherical coordinates $(t,r,\theta,\varphi)$, where $\theta$ and $\varphi$ are the azimuthal and polar angles. By the $O(3)$ symmetry, the Klein-Gordon equation becomes
\be \label{eq:O3evolution}
\partial_t^2 \phi - \partial_r^2 \phi - \frac{2}{r} \partial_r \phi =  -V'(\phi)\,.
\ee
If the initial bubble is $O(4)$ symmetric its evolution in addition to $O(3)$ rotation symmetry invariant also in Lorentz boosts, i.e. it respects $O(1,3)$ symmetry. In this case it is convenient to define new coordinates $(s,\psi)$ by
\bea \label{eq:spsi}
&t = s\cosh\psi \,, \quad r = s\sinh\psi \,, \qquad {\rm for}\,\, t\geq r \,, \\
&t = s\sinh\psi \,, \quad r = s\cosh\psi \,, \qquad {\rm for}\,\, t<r \,.
\eea
The derivative $\partial\phi/\partial\psi$ vanishes, and the Klein-Gordon equation reduces to one dimension,
\be \label{eq:O4evolution}
\pm\partial_s^2 \phi \pm \frac{3}{s} \partial_s \phi =  - V'(\phi)\,,
\ee
where $+$ sign corresponds to $t\geq r$ and $-$ sign to $t<r$. We see that in the region $t<r$ we reproduce Eq.~\eqref{eq:O4initialbubble} and therefore the solution is simply $\phi(s) = \phi_0(s)$, or $\phi(t,r) = \phi_0(\sqrt{r^2-t^2})$, where $\phi_0$ denotes the solution of~\eqref{eq:O4initialbubble}. In particular, the bubble wall traces the hyperboloid $r_0^2 = r^2 - t^2$.

Let us next consider collision of two bubbles. We arrange the coordinate system such that the bubble centers lie on the $z$ axis at $z=\pm d/2$. A collision of two $O(3)$ symmetric bubbles is invariant under $O(2)$ group consisting of rotations in the $(x,y)$ -plane perpendicular to the collision axis $z$. The evolution of this two-bubble system is most easily described in cylindrical coordinates $(t,r,\theta,z)$ in which the Klein-Gordon equation simplifies due to the $O(2)$ symmetry to
\be
\partial_t^2 \phi - \partial_r^2 \phi - \frac{1}{r} \partial_r \phi - \partial_z^2 \phi = -V'(\phi) \,.
\ee
If the colliding bubbles are instead $O(1,3)$ symmetric, their collision is $O(1,2)$ symmetric, and the Klein-Gordon equation is given by
\be
\pm\partial_s^2 \phi \pm \frac{2}{s} \partial_s \phi - \partial_z^2 \phi = -V'(\phi) \,,
\ee
where, again, $+$ and $-$ signs correspond to the regions $t\geq r$ and $t<r$, respectively. The bubble collision happens in the region $t\geq r$. In this region we solve the above equation numerically, as described in the next section. In the region $t<r$ we can, instead, simply use the analytical continuation of the initial bubble solution, 
\bea
\phi(s,z) =& \phi_0\left[\sqrt{s^2+(z-d/2)^2}\right] \\&+ \phi_0\left[\sqrt{s^2+(z+d/2)^2}\right] \,,
\eea
where $s^2 = -t^2 + r^2$.

\section{Two-bubble collisions}
\label{sec:simulations}

\begin{figure}
\centering
\includegraphics[width=\columnwidth]{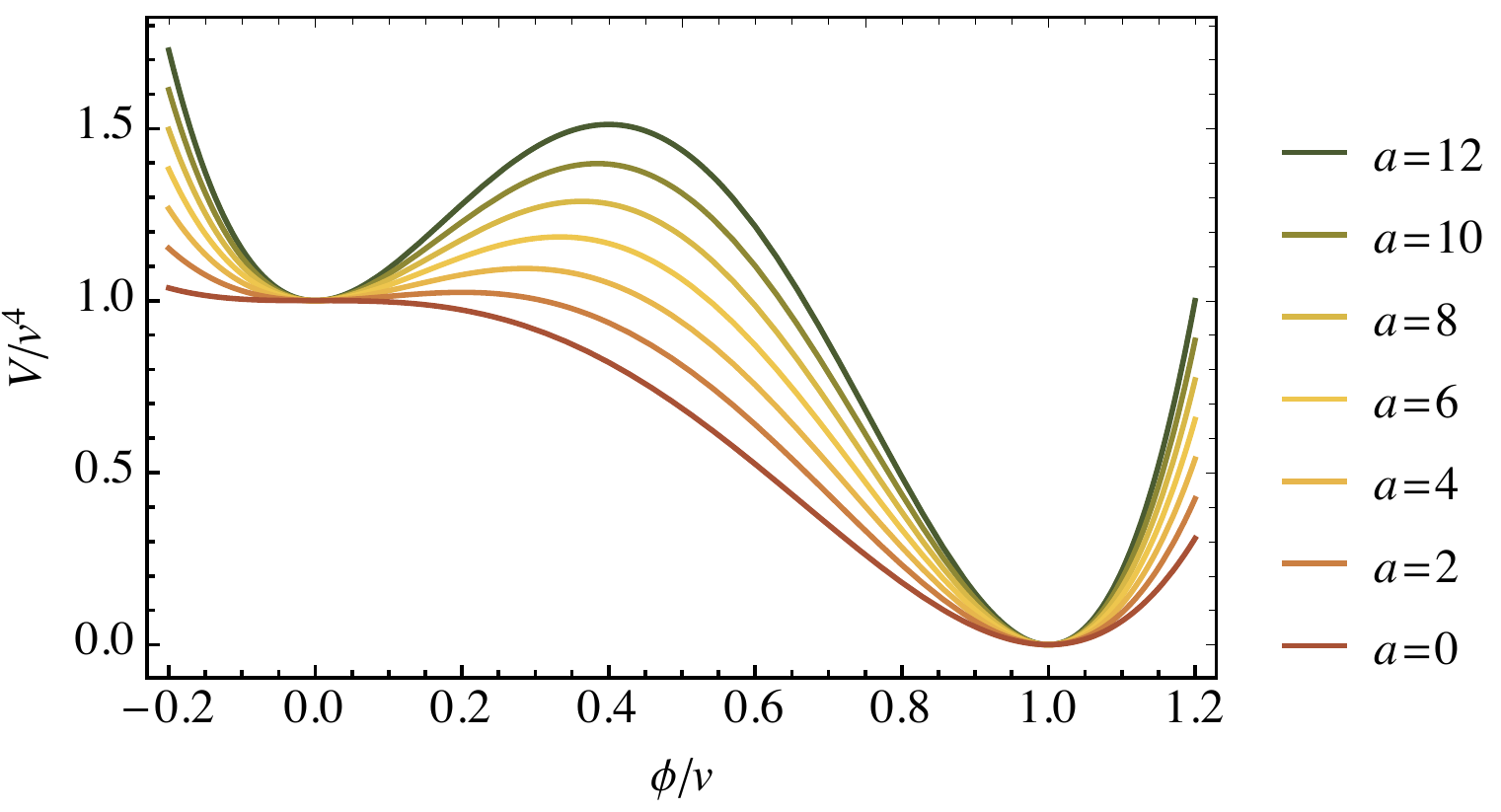}
\caption{The scalar potential~\eqref{eq:pot} used in the numerical simulations for different values of the parameter $a$.}
\label{fig:pot}
\end{figure}

Next we turn to discuss the numerical simulations of two-bubble collisions. For this, we need to pick a scalar potential. Due to its simplicity, we consider a polynomial scalar potential with two minima, given by
\be \label{eq:pot}
V(\phi) = v^4 + av^2\phi^2 - (2a+4)v\phi^3 + (a+3)\phi^4 \,,
\ee
where $a,v>0$. The global minimum of this potential is at $\phi = v$ with $V(v)=0$, and $\phi=0$ is a local minimum with $V(0) = v^4$. The parameter $a$ determines the position and height of the barrier between the minima, as illustrated in Fig.~\ref{fig:pot}. For $a=0$ the barrier vanishes. In the following we use units where $v=1$.


We start from the state where the field $\phi$ lies in the false vacuum at $\phi=0$. At $t=0$ we put 
two identical bubbles at $z=\pm d/2$, and choose their distance $d$ to be much larger than their initial radius. We consider both $O(3)$ and $O(4)$ symmetric initial bubbles calculated by solving Eq.~\eqref{eq:O3initialbubble} in the $O(3)$ symmetric case and Eq.~\eqref{eq:O4initialbubble} in the $O(4)$ symmetric case. We note that in both cases the initial profile can be well fit by a tanh -profile. Moreover, in the $O(3)$ case we need to start from a bubble that is slightly bigger than the solution of Eq.~\eqref{eq:O3initialbubble} as the exact solution would remain stationary. 

We then evolve the scalar field by solving numerically on a lattice Eq.~\eqref{eq:O3evolution} in the case of $O(3)$ symmetric initial bubbles and Eq.~\eqref{eq:O4evolution} in the case of $O(4)$ symmetric initial bubbles. The simulation in the latter case is slightly simpler, as it includes one coordinate less (only $s$ and $z$) than in the $O(3)$ symmetric case ($t$, $r$ and $z$). We consider only identical bubble collisions, so we implement reflecting boundary conditions at $z=0$ and simulate only the region $z\geq 0$.

\begin{figure}[b]
\centering
\hspace{-3mm}\includegraphics[width=\columnwidth]{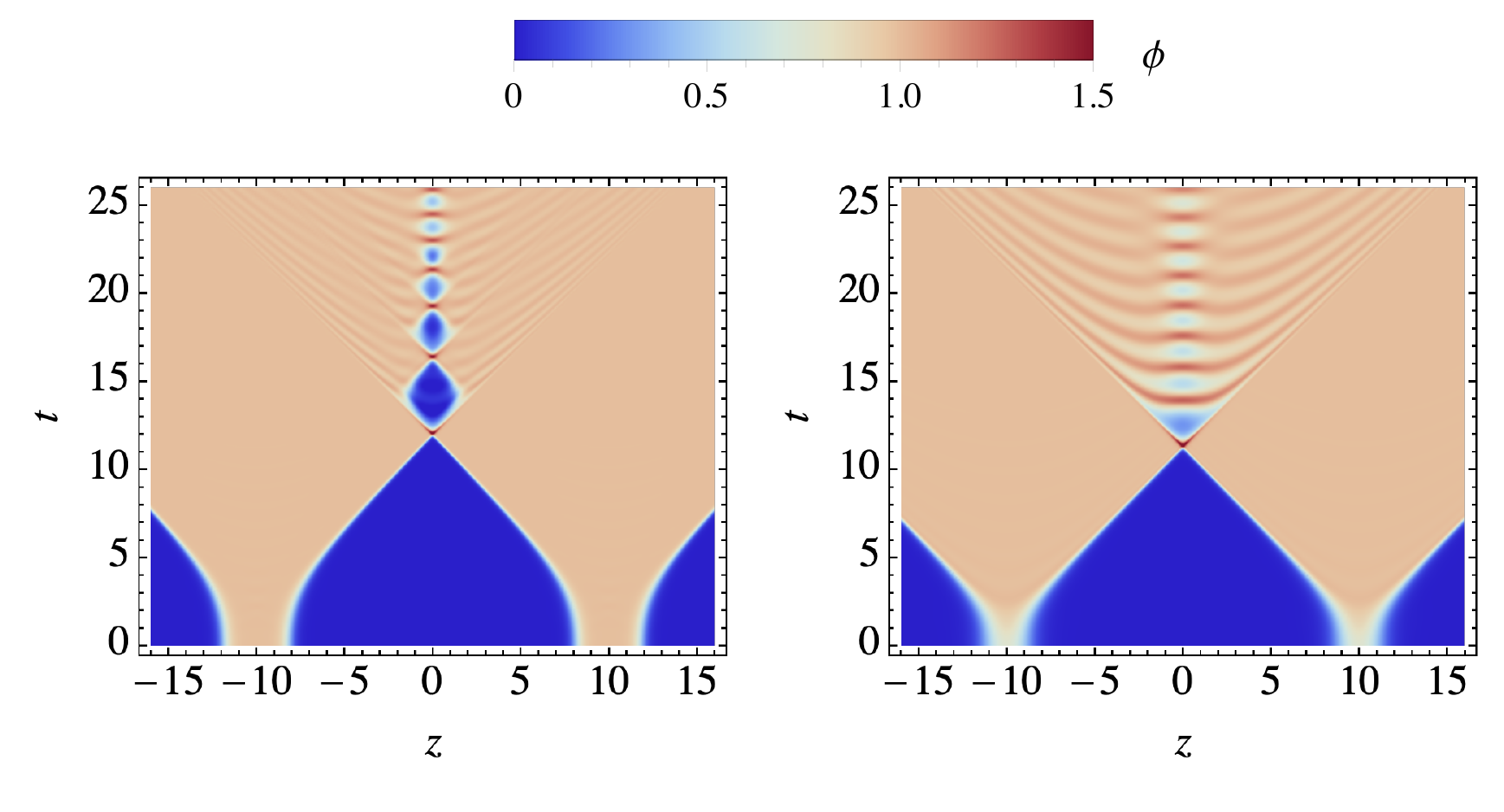} \\ \vspace{-2mm}
\hspace{-3mm}\includegraphics[width=\columnwidth]{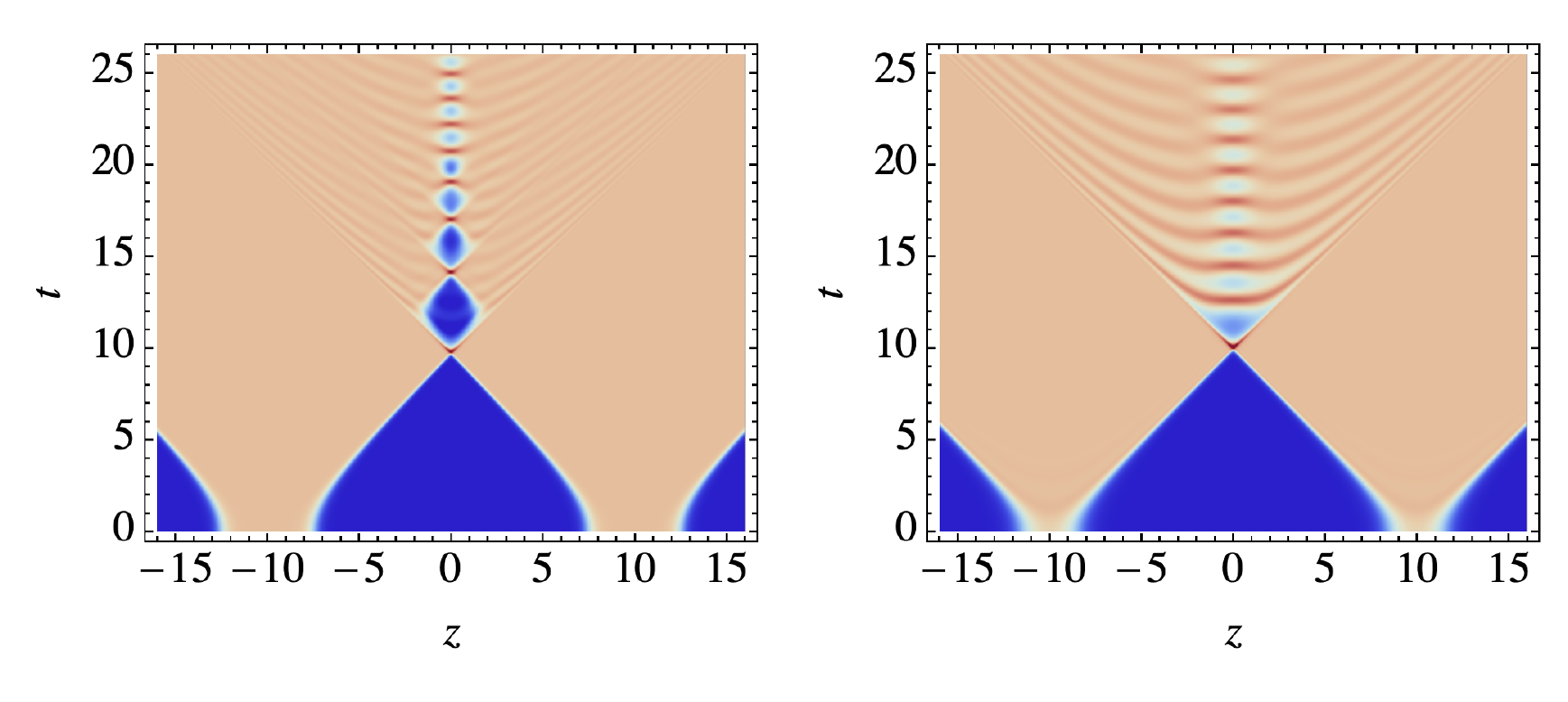} \\ \vspace{-4mm}
\caption{The time evolution of the scalar field at $r=0$ in collision of two identical bubbles. The upper and lower panels correspond to $O(3)$ and $O(4)$ symmetric initial bubbles, and the left and right panels to $a=10$ and $a=2$.}
\label{fig:comparison}
\end{figure}

We first compare the cases with $O(4)$ and $O(3)$ symmetric bubble nucleations. As can be seen in Fig.~\ref{fig:comparison} the results in these cases for the evolution after the bubble collision are almost identical. This is because the radius of the initial bubbles is much smaller than the distance between them, and therefore the initial configuration has only a small effect on the evolution after the collision. This also works as a good check that the simulations work properly. In the following we will focus on the results of simulations with $O(4)$ symmetric initial bubbles as these simulations are faster. However, we have cross-checked all our results in the $O(3)$ case.

\begin{figure}
\centering
\hspace{-3mm}\includegraphics[width=\columnwidth]{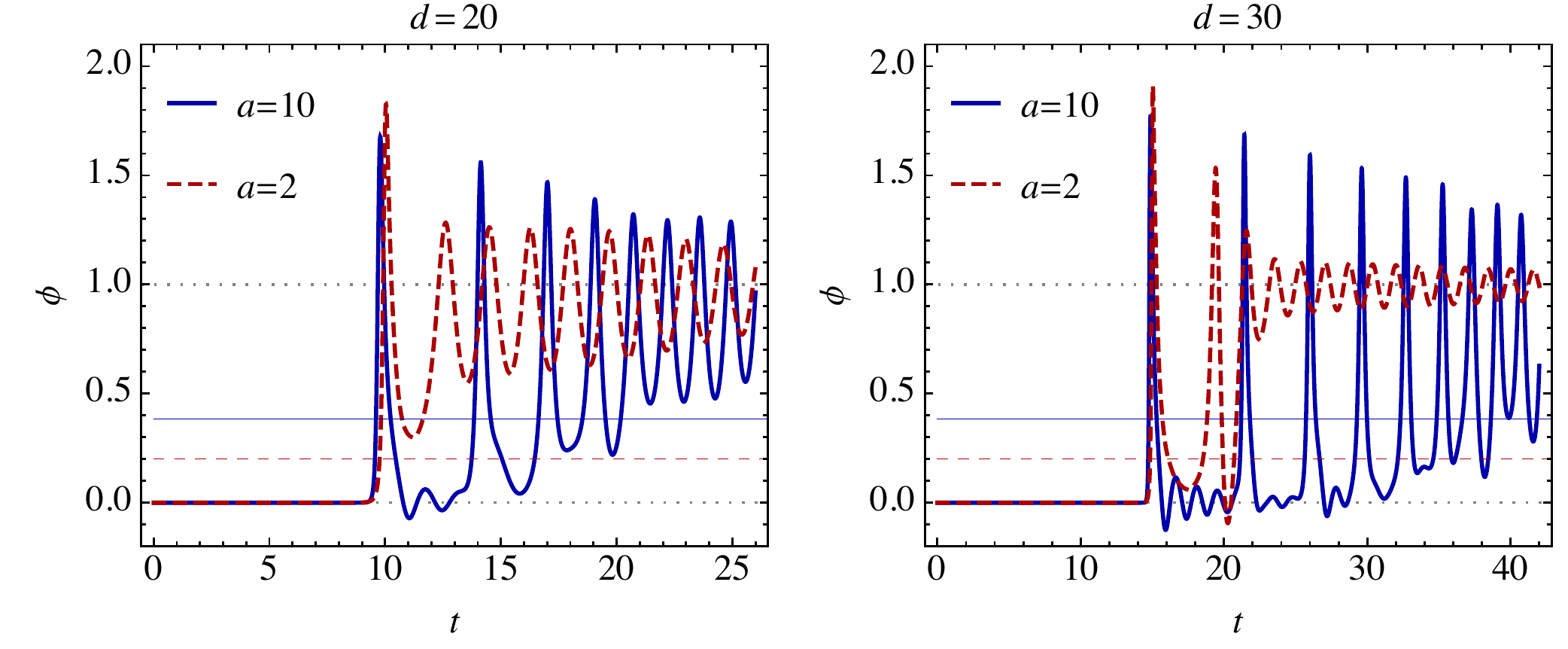}
\\ \vspace{-4mm}
\caption{The time evolution of the scalar field $r=z=0$ in two-bubble collision starting from two identical $O(4)$ symmetric bubbles. The left and right panels correspond to $d=20$ and $d=30$, and the red dashed and blue solid lines to $a=10$ and $a=2$. The thin horizontal lines show the position of the maximum of the scalar potential.}
\label{fig:evol}
\end{figure}

\begin{figure*}
\centering
\includegraphics[width=\textwidth]{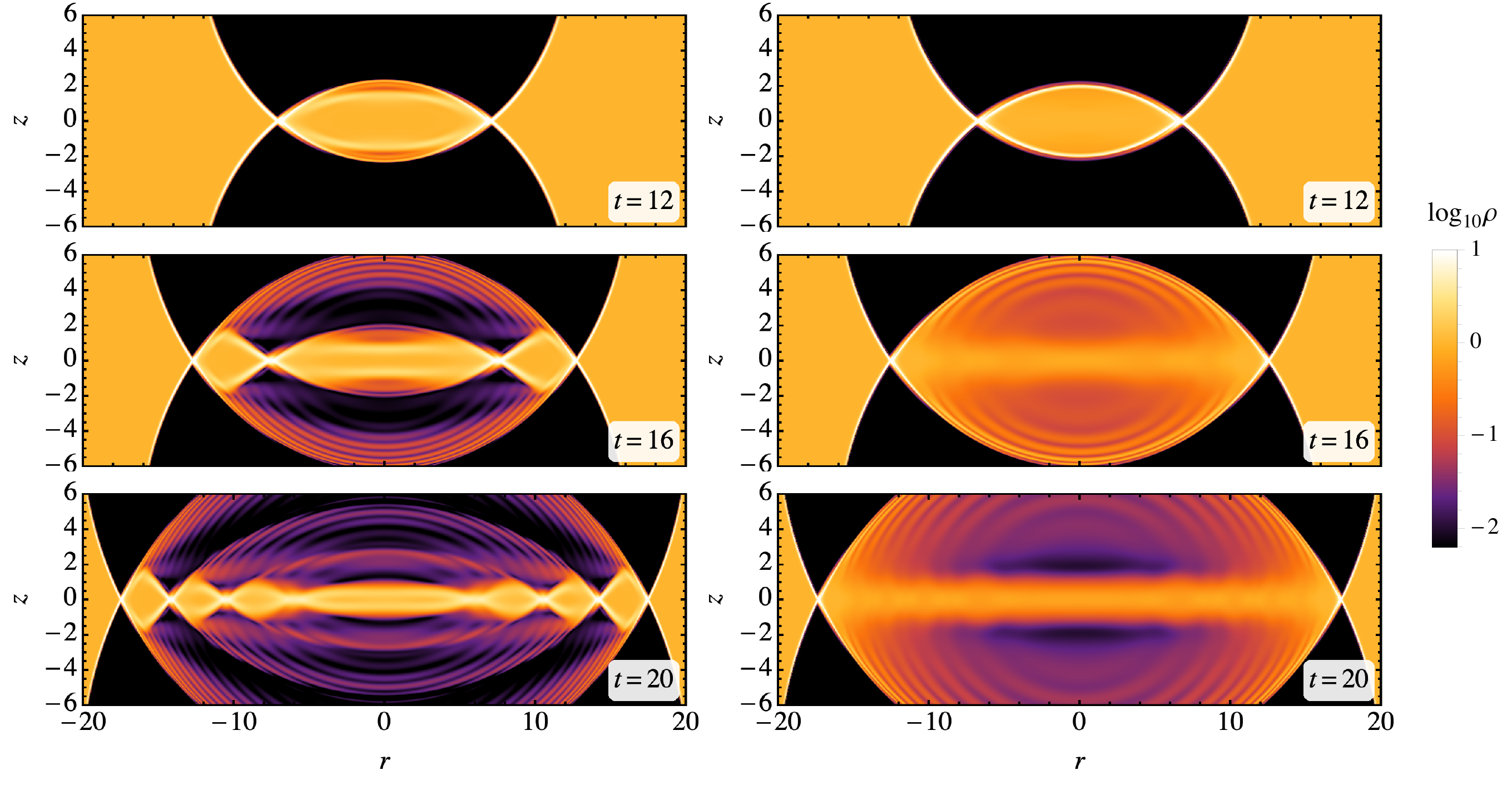}
\vspace{-2mm}
\caption{Snapshots of the scalar field energy density profile in two-bubble collision starting from two identical $O(4)$ symmetric bubbles. The left and right panels correspond to $a=10$ and $a=2$. The bubble centers are separated by distance $d=20$.}
\label{fig:rzplane}
\end{figure*}

From Fig.~\ref{fig:evol}, where we show the field value at the collision point $z=r=0$, we see that in the case $a=10$ the field bounces back to the false vacuum whereas in the case $a=2$ it remains oscillating around the true vacuum. The difference between these cases can be very clearly seen in Fig.~\ref{fig:rzplane}, where we show the energy density of the scalar field. The energy density is on the $r=0$ surface given by
\be
\rho(s,z) = \frac{1}{2} \left(\frac{\partial \phi}{\partial s}\right)^2 +  \frac{1}{2}\left(\frac{\partial \phi}{\partial z}\right)^2 + V(\phi) \,,
\ee
and can be, similarly to the field value, continued to non-zero values of $r\leq t$ by replacement $s\to \sqrt{t^2-r^2-z^2}$. In the former case the energy density after collision is focused dominantly in a small region around the collision plane whereas in the latter case it spreads more uniformly. This may have an effect on the GW signal from the collision as discussed in Sec.~\ref{sec:gws}. Moreover, as discussed in Sec.~\ref{sec:bhs} the latter case has been studied in context of black hole formation.

As described in Ref.~\cite{Hawking:1982ga}, in the case that a false vacuum bubble is formed, the bubble walls effectively bounce from each other in the collision and start slowing down due to the potential energy difference and bubble wall tension. The false vacuum bubble eventually starts to shrink and if it still has enough energy it will bounce again. Eventually it has lost so much energy that the bubble walls have become non-relativistic and they disappear. The size of the false vacuum bubble and how long it bounces before disappearing depends on the distance between the bubbles, as can be seen in Fig.~\ref{fig:evol}, and the potential energy difference between the minima.

In a more complicated scalar potential with more local minima, the field may in the collision bounce to a different false vacuum than the original one, or even to a deeper minimum than the one inside the initial bubbles. In the latter scenario the region around the collision point may then continue growing if the potential energy difference is sufficiently large. This mechanism may bring the system to the true vacuum even if the nucleation probability of the true vacuum bubbles is too small. In this study we will, however, consider only the simplest case where the potential has two minima.

The trapping equation~\cite{Jinno:2019bxw}
\be \label{eq:trapping}
\partial_s^2 \phi + \frac{1}{s} \partial_s \phi =- \frac{\partial V}{\partial\phi}
\ee
can be used to approximately determine whether a false vacuum bubble is formed at the collision point. In this simple description appropriate for infinitely energetic bubbles, at the moment of collision the two field profiles simply add giving the initial value $\phi(s=0)=2\phi_{\rm min}$ for the above equation. For the scalar potential~\eqref{eq:pot} the larger $a$ is the bigger the barrier between the vacua and for sufficiently large $a$ the field will be stuck in the false vacuum after bouncing back there in the collision. Eq.~\eqref{eq:trapping} predicts the false vacuum region is formed in the bubble collisions for $a>7.76$. However, in practice as already pointed out in~\cite{Jinno:2019bxw} simulations never feature truly infinitely energetic bubbles and as a result the dividing line between these cases is more complicated. We will use two examples with $a=10$ and $a=2$. The evolution of the field in these cases is showed in Fig.~\ref{fig:evol} and we see that even though for $d=30$ the field bounces twice to the false vacuum even for $a=2$, it is not trapped there and quickly rolls back to the true vacuum.

\section{Gravitational waves}
\label{sec:gws}

In this section we study the GW signal from a collision of two identical bubbles. For very strong transitions, considered in this paper, the gradients in the scalar field give the main source of GWs. The GW spectrum was in this case first calculated for two-bubble collisions in Ref.~\cite{Kosowsky:1991ua}. These calculations suggested that envelope approximation, where the bubble walls are treated as infinitely thin shells and the collided parts of the bubble walls are neglected, could give a good description of the GW production in bubble collisions~\cite{Kosowsky:1992vn}. This approximation has since been revised in larger simulations~\cite{Huber:2008hg,Weir:2016tov,Jinno:2016vai}, and its result compared to that of full lattice simulations~\cite{Cutting:2018tjt}. 

In this paper our goal is to compare the GW signals from the cases with and without trapping. We numerically calculate the GW signal in two-bubble collisions leaving a more complete treatment with multiple bubbles for future work.

The total energy integrated over directions of the wave vector $\khat$ at frequency $\omega$ of the GWs emitted in the phase transition is given by~\cite{Weinberg:1972kfs}
\be \label{eq:dEkdw}
\frac{\td E}{\td \omega} = 2G\omega^2 \int\td\Omega_k\, \Lambda_{ijlm}(\khat) T_{ij}^*(\khat,\omega) T_{lm}(\khat,\omega) \,,
\ee
where $\Lambda_{ijlm}$ is the projection tensor,
\bea
\Lambda_{ijlm}(\khat) =& \,\delta_{il}\delta_{jm} -2\delta_{il}\khat_j\khat_m + \frac{1}{2}\khat_i \khat_j \khat_l \khat_m \\ 
&- \frac{1}{2} \delta_{ij}\delta_{lm} + \frac{1}{2}\delta_{ij}\khat_l\khat_m + \frac{1}{2}\delta_{lm}\khat_i\khat_j \,,
\eea
and the traceless part of the stress energy tensor sourced by the scalar field gradients is
\be
T_{ij}(\khat,\omega) = \frac{1}{2\pi} \int \td t\, \td^3 x \,e^{i\omega(t-\khat\cdot\boldsymbol{x})} \,\partial_i \phi \partial_j \phi \,.
\ee

We consider two-bubble collisions and set the collision point as the origin of the coordinate system and $z$ axis along the symmetry axis of the system. The integral over the azimuthal angle can be performed analytically as the derivatives don't depend on it. Moreover, we can take without loss of generality $\khat = (\sin\theta_k,0,\cos\theta_k)$, which implies that the $xy$ and $yz$ components of $T_{ij}$ vanish. The remaining components of $T_{ij}$ can be written as
\begin{widetext}
\bea
&T_{rr}(\theta_k,\omega) = -\frac{1}{2}\int\! \td t\,e^{i\omega t} \int_0^\infty\!\td z\cos(\omega \cos\theta_k z)\!\int_0^\infty\!\td r \,r \left[\sin^2\theta_k J_0(\omega \sin\theta_k r) + (\cos^2\theta_k\!+\!1) J_2(\omega \sin\theta_k r)\right] (\partial_r \phi)^2 \,, \\
&T_{xz}(\theta_k,\omega) = -\int \td t\,e^{i\omega t} \int_0^\infty\!\td z\,\sin(\omega \cos\theta_k z) \int_0^\infty\!\td r\,r \,J_1(\omega \sin\theta_k r)\, \partial_r \phi\, \partial_z \phi\,, \\
&T_{zz}(\theta_k,\omega) = \int \td t\,e^{i\omega t} \int_0^\infty\!\td z\,\cos(\omega \cos\theta_k z) \int_0^\infty\!\td r\,r \,J_0(\omega \sin\theta_k r)\, (\partial_z \phi)^2 \,.
\eea 
where $T_{rr}(\theta_k,\omega) \equiv \cos^2\theta_k T_{xx}(\theta_k,\omega) - T_{yy}(\theta_k,\omega)$ and $J_j$ denote Bessel functions of the first kind. The integral over $\khat$ directions in Eq.~\eqref{eq:dEkdw} reduces to $\td \Omega_k \to 2\pi \td \theta_k \sin\theta_k$, and the projection gives
\be
\frac{\td E}{\td \omega} = 2\pi G\omega^2 \int_0^\pi \td \theta_k \sin\theta_k \left|T_{rr}(\theta_k,\omega) + \sin^2\theta_k T_{zz}(\theta_k,\omega) - 2\sin\theta_k\cos\theta_k T_{xz}(\theta_k,\omega)\right|^2 \,,
\ee  

In the $O(1,2)$ symmetric case we use the $(s,\psi)$ coordinates defined in Eq.~\eqref{eq:spsi}. Let us denote by $\phi_\pm$ the solution at $t\geq r$ and $t<r$, respectively. The $r$ derivative is then given by $\partial_r \phi_+ = -\sinh\psi \partial_s \phi_+$ for $t\geq r$ and $\partial_r \phi_- = \cosh\psi \partial_s \phi_-$ for $t<r$, and integrals can be expressed as
\bea
&
\begin{split}
T_{rr}(\theta_k,\omega) = -&\frac{1}{2}\int \td s\, s^2 \int_0^\infty\!\td z\,\cos(\omega \cos\theta_k z) \\
&\times \bigg[(\partial_s \phi_+)^2 \int_0^\infty\!\td\psi\sinh^3\psi \,e^{i\omega s \cosh\psi} \!\left[\sin^2\theta_k J_0(\omega \sin\theta_k s \sinh\psi) + (\cos^2\theta_k\!+\!1) J_2(\omega \sin\theta_k s \sinh\psi)\right]  
\\ &\,\,\,\,+ (\partial_s \phi_-)^2 \int_0^\infty\!\td\psi\cosh^3\psi \,e^{i\omega s \sinh\psi} \!\left[\sin^2\theta_k J_0(\omega \sin\theta_k s \cosh\psi) + (\cos^2\theta_k\!+\!1) J_2(\omega \sin\theta_k s \cosh\psi)\right]\bigg] \,, 
\end{split}
\\ &
\begin{split}
T_{xz}(\theta_k,\omega) = \int \td s\, s^2 \int_0^\infty\!\td z\,\sin(\omega \cos\theta_k z) \bigg[& \partial_s \phi_+ \partial_z \phi_+ \int_0^\infty\!\td \psi\,\sinh^2\psi \,e^{i\omega s \cosh\psi} J_1(\omega \sin\theta_k s \sinh\psi) 
\\ &- \partial_s \phi_- \partial_z \phi_- \int_0^\infty\!\td \psi\,\cosh^2\psi \,e^{i\omega s \sinh\psi} J_1(\omega \sin\theta_k s \cosh\psi) \bigg] \,,
\end{split}
\\ &
\begin{split}
T_{zz}(\theta_k,\omega) = \int \td s\, s^2 \int_0^\infty\!\td z\,\cos(\omega \cos\theta_k z) \bigg[&  (\partial_z \phi_+)^2\int_0^\infty\!\td \psi\,\sinh\psi \,e^{i\omega s \cosh\psi} J_0(\omega \sin\theta_k s \sinh\psi)
\\ &+  (\partial_z \phi_-)^2 \int_0^\infty\!\td \psi\,\cosh\psi \,e^{i\omega s \sinh\psi} J_0(\omega \sin\theta_k s \cosh\psi) \bigg] \,.
\end{split}
\eea
\end{widetext}
We note that the field derivatives are independent of $\psi$, and for fixed values of $s$, $\theta_k$ and $\omega$, we can perform the $z$ and $\psi$ integrals separately. This makes the numerical calculation in the case of $O(4)$ symmetric initial bubbles significantly faster than the one with $O(3)$ symmetric initial bubbles.

\begin{figure*}
\centering
\includegraphics[width=0.48\textwidth]{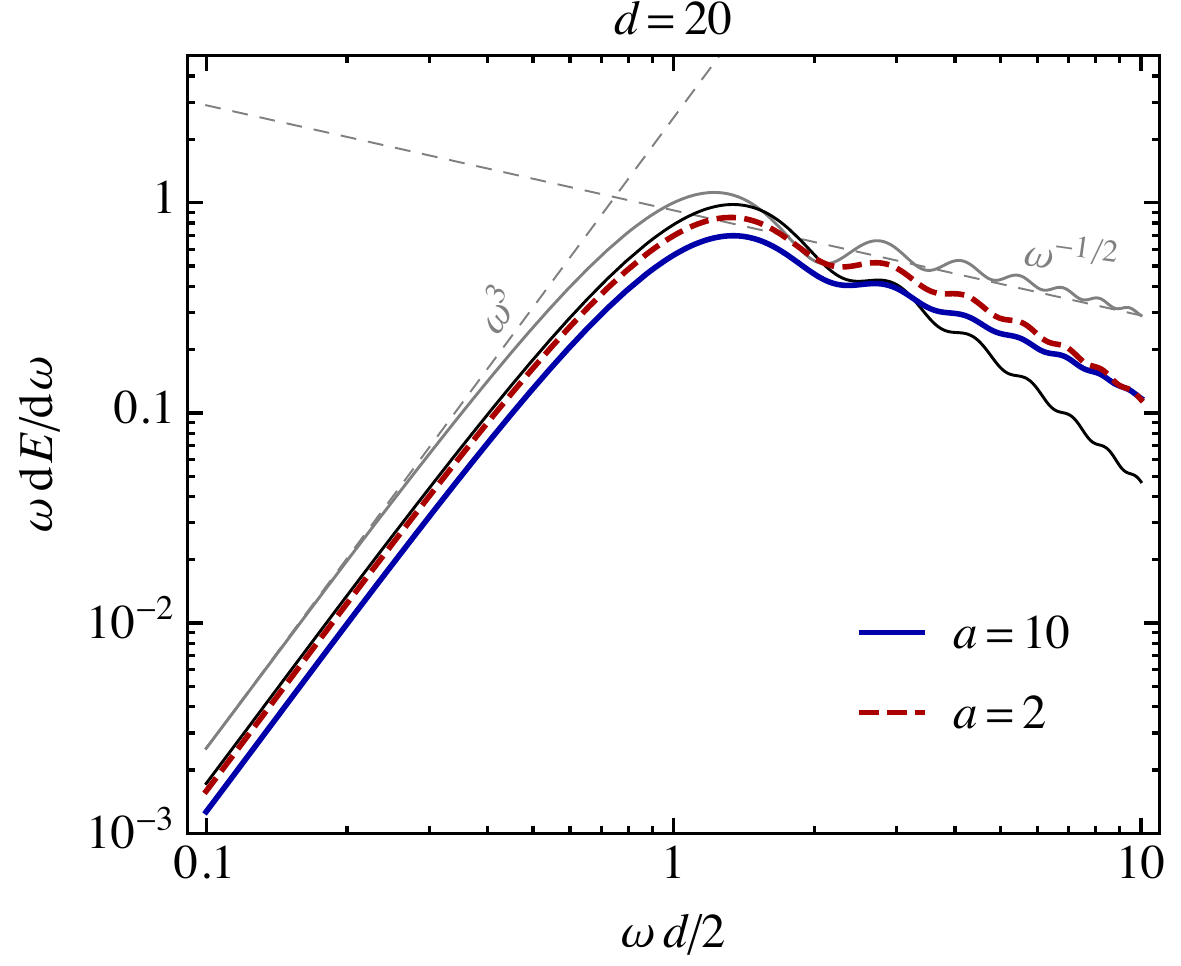} 
\hspace{3mm}
\includegraphics[width=0.48\textwidth]{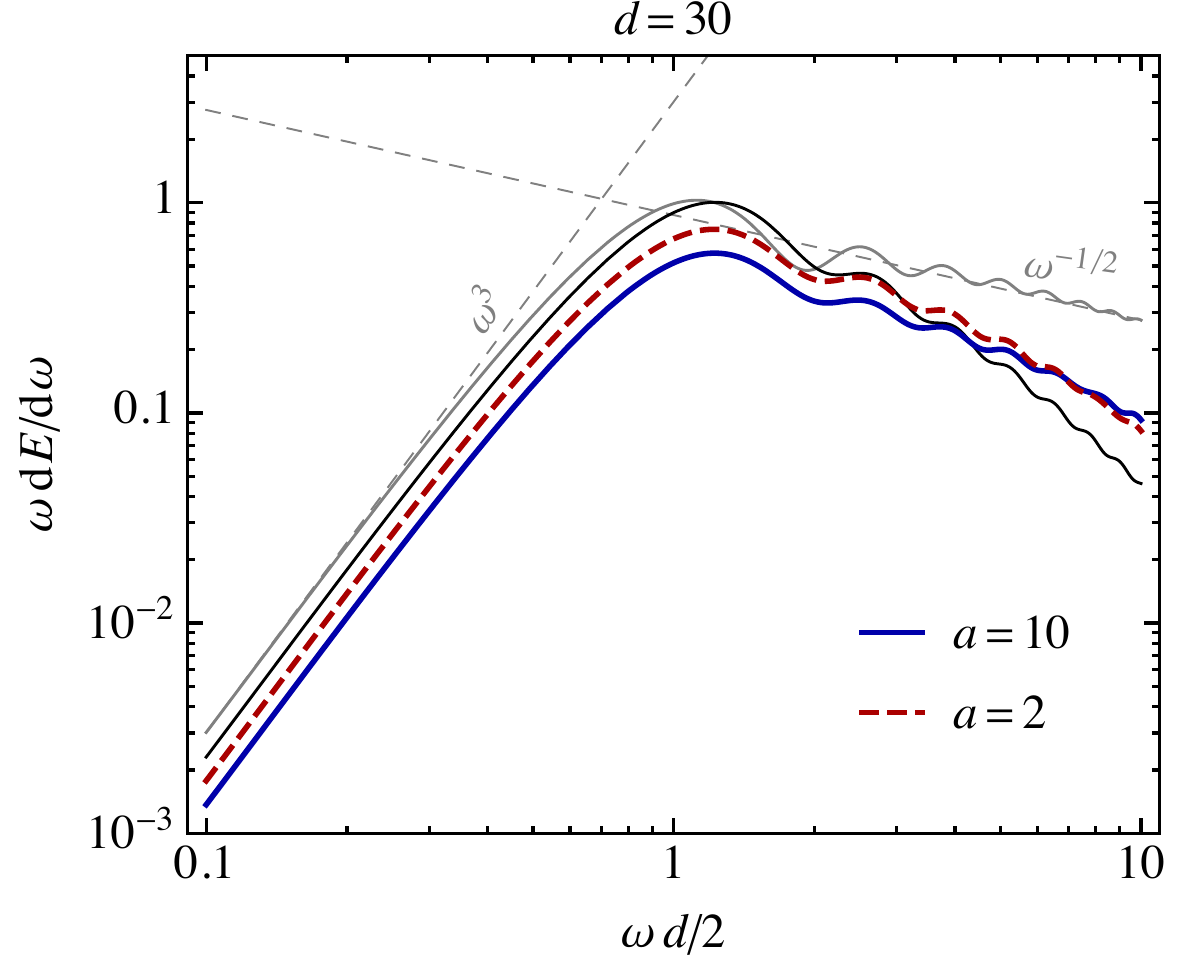}
\caption{GW spectrum from two-bubble collision. The red dashed curve shows the case where field remains oscillating around the true vacuum after the collision and the blue solid curve the case where it bounces to the false vacuum. For comparison, the envelope approximation result is shown by the thin gray line and the bulk flow approximation by the thin black line. The left and right panels correspond to different distances $d$ between the bubbles. All curves are divided by the maximum amplitude of the envelope approximation result.}
\label{fig:gws}
\end{figure*}

The result from the simulations with $O(4)$ symmetric initial bubbles are shown in Fig.~\ref{fig:gws}. The blue solid and red dashed curves correspond to the GW spectra in the cases with and without formation of false vacuum bubbles in the collision. We show the results for two different distances between the bubble centers. For $d=20$ the simulation ends at $t_{\rm max}=26$ and for $d=30$ at $t_{\rm max}=42$. We have not included a smooth end for the GW calculation as was done in Ref.~\cite{Kosowsky:1991ua}. However, we have checked that the shape of the GW spectrum is not sensitive to the value of $t_{\rm max}$ which rather only changes the overall amplitude of the spectrum.

We see that the peak amplitude is higher in the case without formation of false vacuum bubble in the collision. This is simply because the initial bubble is in that case smaller (see e.g. Fig.~\ref{fig:evol}) and the bubble wall therefore reaches a higher velocity before the collision leading to production larger GW energy density. The Lorentz factors of the bubble walls at the collision are $\gamma_w \approx 4$ for $a=10$ and $\gamma_w \approx 7$ for $a=2$ in the case $d=20$, and $\gamma_w \approx 6$ for $a=10$ and $\gamma_w \approx 10$ for $a=2$ in the case $d=30$. 

Moreover, it seems that the spectrum drops at high frequencies faster than in the case where a false vacuum bubble is formed in the collision and this difference seems to increase as a function of $d$. However, the difference is quite small and from this result it seems unlikely that these cases can be separated by their GW signal. However, in the case of more than two bubbles the spectrum is affected also by the distribution of bubble sizes (or nucleation times), and the fact that the bubble wall energy propagates for a larger distance from the collision point in the case without false vacuum bubble formation may change the spectrum similarly as in the bulk flow approximation introduced in Ref.~\cite{Konstandin:2017sat}.

For comparison, we show also the result obtained in the envelope approximation by the gray line~\cite{Kosowsky:1992vn} and in the bulk flow approximation by the black line~\cite{Konstandin:2017sat}. Whereas in the envelope approximation the collides parts of the bubbles are completely neglected, in the bulk flow approximation the collided sections lose their energy gradually. The amplitude in the envelope approximation is slightly higher than from the full calculation but it predicts very well the position of the peak. All results give the same slope, $\sim \omega^3$, at small frequencies, which simply arises from the fact that the stress energy tensor $T_{\rm ij}$ asymptotically reaches a constant value at $\omega\ll d/2$. At high frequencies the spectrum in the envelope approximation drops significantly slower than the result of the full calculation. The envelope approximation gives slope $\sim \omega^{-1/2}$ at high frequencies whereas the slope of the result of full calculation is between $\sim\omega^{-1}$ and $\sim\omega^{-3/4}$. In the bulk flow approximation the spectrum instead drops too fast at high frequencies, with slope between $\sim \omega^{-1}$ and $\sim \omega^{-2}$ in the range $\omega d/2 < 10$.

\section{Black hole formation}
\label{sec:bhs}

Despite extensive experimental efforts the non-gravitational nature of dark matter (DM) is largely unknown. It can consist of almost anything from very light axion like particles to macroscopic primordial black holes (PBHs). As the WIMP paradigm, that is based on the formation of the observed DM abundance through freeze-out mechanism, has become strictly constrained by the non-observation of WIMPs in direct DM searches~\cite{Escudero:2016gzx,Arcadi:2017kky}, alternative scenarios have become increasingly popular. These include for example formation of the observed DM abundance through freeze-in mechanism~\cite{Hall:2009bx,Bernal:2017kxu} or vacuum realignment~\cite{Kim:1986ax,Turner:1989vc}. In these scenarios the DM can be hidden from the direct searches as it's couplings to Standard Model (SM) particles are tiny. 

Also the possibility that the DM is in PBHs that are too heavy to have evaporated by now has recently been revived, and various constraints on their abundance have been revised. It has been shown that all DM can be in asteroid mass PBHs~\cite{Carr:2017jsz,Katz:2018zrn}. Moreover, many different models for their formation have been thoroughly studied~\cite{GarciaBellido:1996qt,Kawasaki:2016pql,Kawasaki:1997ju,Drees:2011hb,Drees:2011yz,Kawasaki:2012wr,Clesse:2015wea,Garcia-Bellido:2017mdw,Ezquiaga:2017fvi,Kannike:2017bxn,Dimopoulos:2019wew}. These are dominantly based on generation of peaks in the curvature power spectrum during inflation which eventually collapse to BHs as they re-enter horizon after inflation~\cite{Carr:1974nx,Carr:1975qj}. Other PBH formation scenarios include collapse of cosmic strings~\cite{Garriga:1992nm,Caldwell:1995fu}, domain walls~\cite{Garriga:1992nm,Deng:2016vzb}, or string-domain wall network~\cite{Ferrer:2018uiu}, scalar field fragmentation~\cite{Cotner:2016cvr,Cotner:2018vug,Cotner:2019ykd}, and nucleation of false vacuum bubbles during inflation~\cite{Garriga:2015fdk,Deng:2017uwc,Kusenko:2020pcg}. 

In strongly supercooled phase transitions where the expanding bubbles and their collisions generate large overdensities, as can be seen in Figs.~\ref{fig:evol} and \ref{fig:rzplane}. Therefore a natural question to ask is if BHs are formed in bubble collisions. The first attempt to do this based on the overdensities caused by multiple bubble walls intersecting at one point~\cite{Hawking:1982ga}. However, the volumes containing enough energy for gravitational collapse are much larger than the intersection point and the system is then very far from being spherically symmetric. It is therefore not clear if a BH is formed. 

Another proposed mechanism relies on gravitational collapse of small regions of false vacuum existing still just before the transition completes~\cite{Kodama:1982sf}. Such final false vacuum regions left between growing bubbles are typically not spherically symmetric. In fact, it is expected that the energy contained in such a region drops quickly as it shrinks. This, however, does not mean that a BH can not be formed, but a dedicated numerical study is necessary to study the evolution of the last false vacuum regions and possible density of BHs that could potentially be produced.

Third BH formation mechanism in phase transitions was introduced in Ref.~\cite{Khlopov:1998nm}. It is based on formation of false vacuum bubbles in bubble collision. It was assumed that these false vacuum bubble eventually become spherical, shrink and collapse to a BHs. This seems at first sight a promising formation scenario as it leads to large abdundance or relatively big PBHs. However, our results indicate that the assumption the main assumption behind this mechanism is not realised. From Fig.~\ref{fig:rzplane} we see that, instead of getting spherical, the false vacuum bubble becomes more an more pancakey as it bounces. Therefore, it seems that black hole formation mechanism considered in Ref.~\cite{Khlopov:1998nm} does not work.

\section{Conclusions}

In this paper we considered strongly supercooled phase transitions which provide the most optimistic situation for generating a strong GW background. We studied two-bubble collisions by  numerical lattice simulations. We showed that in the bubble collision the scalar field can either bounce to a false vacuum or remain oscillating around the true vacuum. We then studied if the GW signal from the bubble collision can be used to distinguish these two cases. We found that the produced GW spectra are only slightly different and therefore seems to not allow observing the false vacuum bubble formation. We also compared the GW signal from the full calculation to the results in the envelope and bulk flow approximations, and found considerable differences in the spectrum at high frequencies. We finally discussed the possibility of black hole formation in bubble collisions. Our results indicate that vacuum bubbles do not collapse to black holes as described in Ref.~\cite{Khlopov:1998nm}. \vspace{4mm}

\acknowledgments 
We thank Miguel Escudero, Kimmo Kainulainen for helpful discussions. We would also like to express our gratitude to Nordita and CERN Theory Division who hosted us during our work on this project. This work was supported by the UK STFC Grant ST/P000258/1. ML was partly supported by the Polish National Science Center grant 2018/31/D/ST2/02048 and VV by the Estonian Research Council grant PRG803.

\bibliography{PBH}
\end{document}